\documentstyle[12pt]{article}
\include{epsf}
\include{psfig}
\input{psfig.sty}
\begin{document}

\title{Mathematical modeling of filamentous
microorganisms}
 
\author{Michele Bezzi$^{*}$ and Andrea Ciliberto$^{\dag}$}
\date{$^{*}$ SONY CSL, 6, rue Amyot, 75005, Paris, France.\\ \vspace{.2cm}
$^{\dag}$  Dept. Biology, Virginia
Polytechnic Institute,\\
 Blacksburg, VA, 24061, USA.}
        
\maketitle
\begin{abstract}
Growth patterns generated by filamentous organisms (e.g. actinomycetes 
and fungi) involve  spatial and temporal dynamics at different length scales.
Several mathematical models have been proposed in the last thirty years
to address these specific dynamics.
Phenomenological macroscopic models are able to reproduce the 
temporal  dynamics of colony-related quantities (e.g. colony
growth rate) but do not explain  the development of mycelial
morphologies nor the single hyphal growth.  Reaction-diffusion models are a 
bridge  between macroscopic and microscopic worlds as they produce
mean-field approximations of single-cell behaviors. Microscopic models
describe intracellular events, such as branching, septation and 
translocation. Finally,
completely discrete models, cellular automata, simulate the
microscopic interaction among cells to 
reproduce emergent cooperative behaviors of large colonies.
In this comment, we review a selection of models for each
of these length scales, stressing their advantages and shortcomings.
\end{abstract}

\section{Introduction}

The panorama of mathematical models 
of filamentous microorganisms is quite large, for
several reasons. To begin with,
filamentous organisms grown on solid media, give rise to morphologies (circular colonies with
branched mycelia) that are easily observable but need to be put in
the rigorous frame of mathematics to be properly understood (see~\cite{ben94,ben95,mats93} for
examples of mathematical models of microorganisms growth patterns).
 Qualitative arguments that have proved very useful in other areas of biology simply
could not match the dynamical and spatially extended problems introduced
by mycelial growth. A second reason of interest concerns industrial
applications of filamentous organisms, since they are among the main
producers of enzymes and antibiotics (for a recent review on fungal morphology 
and industrial application see~\cite{Mcintyre2001}).
Not surprisingly, several mathematical
models have been produced to improve growth and catabolyte production
of fungi and actinomycetes. Finally, multicellular life-style of bacteria
has recently attracted much attention, and particularly developmental
processes related with multicellularity in prokaryotes~\cite{shapiro97}.
Among bacteria, {\it Streptomocyes} is one of the most striking examples
of a multicellular bacteria, and differentiation is a well established
and documented process in this genus. 

In this review, we will not cover extensively the published models
of growth of filamentous organisms,
rather we will
discuss a few examples of models at different organization level, 
ranging from the single cell to the whole colony level. 
Modeling mycelial growth can  be undertaken at different 
levels of complexity and length scale. {\it Microscopic} models are focused on single cell scale, 
so they deal with tip growth, branching and  septation (see
for example ~\cite{Christiansen98,Lazlo93}), while
 {\it macroscopic} models study the evolution
of  quantities on the scale of the whole colony, such as
biomass  or total hyphal length (see for example~\cite{Trinci74,Davidson98,Davidson00}).
\footnote{Notice that ``microscopic'' and ``macroscopic'' 
in this context
 do not always correspond to the same terms  used in 
 other biological or physical modeling contexts, 
for instance in our case microscopic means the level of
a single cell or a single hypha, which is in fact
a macroscopic physical object itself,
with its own complex dynamics.}

In principle, microscopic, single-cell description can be applied individually to 
an huge number of cells, to simulate
the macroscopic behavior of the whole colony. 
Unfortunately,  this approach presents many shortcomings: first the 
many non-linearities present
in these models require large scale simulations which can be
computationally inefficient and unfeasible even using modern computers. 
In addition, a big number of equations implies a big number of  
parameters, which are often difficult
to estimate from experiments. The uncertainty of their values 
results in many different possible outcomes in simulations. 
Finally the results are often difficult to understand and 
analyze, so that even  relatively simple phenomena can be obscured 
by the complexity of the model.

A possible link between microscopic and macroscopic approaches consists
in models whose
variables are local  densities 
of hyphae and concentrations of growth-related
substances.
In these models, which can be viewed as a local mean-field approximation
of microscopic models, single-cell individuality and local fluctuations in the number of
cells or molecules are lost, nevertheless  some spatial structure is retained
so that they are able to reproduce some of 
the macroscopic growth patterns of the colony.
Mathematically, 
these are {\it reaction-diffusion models}~\cite{Edelstein82,Edelstein83,Edelstein89,Yang91,Boswell02},
composed of a set of 
coupled partial differential equations with reaction and diffusion terms. 
Reaction-diffusion models have been widely used, especially to
describe pattern-formation.

In the following sections, 
 we first will give a short summary of the main biological
features of growth of filamentous microorganisms (Section~\ref{sec:biology}).
Then, we will present examples of  macroscopic, microscopic,
reaction-diffusion and completely discrete (cellular automata)
models for mycelial growth in
fungal  and bacterial colonies (Section~\ref{sec:Fungi-model}),
stressing the advantages and limitations of the different approaches.
Finally, in Section ~\ref{sec:aerial}, we will discuss
pattern formation for aerial mycelium.
General conclusions are drawn in the last section.

\section{Growth of filamentous microorganisms}
\label{sec:biology}

In bacteria that divide by binary cell division,
mother and daughter cells separate at the end of the
cell replication process, with a 
minimal spatial displacement: they basically pile up on each other. 
Cell colony increases in size, and exploits the nearest 
resources, until they have been depleted.
The limits of this strategy are particularly evident for growth on
solid (typically agar) media, where colonies are quite small and stop growth
after a few days. Particularly, these cells
are neither able to escape from a region of low nutrient concentration,
nor able to avoid depletion of resources by restricting growth
to a few specialized cells. In the contrary,  a filamentous prokaryote such as {\it Streptomyces}
is able to colonize a whole agar plate, even in minimal medium.
How is that possible?

In filamentous organisms, cells are organized in hyphae, filaments
of cells divided by septa (for reviews on filamentous organisms
growth and life cycle see Refs.~\cite{Prosser91,Trinci94}). At cell division, mother and daughter cells
are not physically separated, but are kept together by the septum,
Fig.~\ref{Fig:hyphal_growth}. This physical constraint has wide
effects on the mechanisms of growth and division: are mother and daughter
cells identical concerning growth and division? The answer to this question lies
at the core of the strategies implemented by filamentous organisms
to exploit environmental resources, and consequently to the growth
kinetics and patterns observed on solid media. 

In this section, we set the stage to answer this key question, 
describing the basic mechanisms of cell division
and hyphal growth, and we explore the most common morphologies and
growth kinetics of filamentous organisms 
on solid media.

\subsection{Hyphal growth}

A typical colony of fungi or actinomycetes grown on solid media is formed by two
different classes of mycelia: vegetative and aerial mycelium.
The former penetrates the medium, and is responsible for
absorbing nutrients from the environment, while the latter, developed
from the vegetative mycelium, releases spores in the air.

We  shall describe hyphal growth and kinetics following the development of a colony,
from germination to the formation of a fully developed colony.
A rigorous general description of
filamentous organisms is impossible since growth mechanisms differ
between fungi and actinomycetes, and even among one actinomycete (or fungus) species to
another.  However, it is possible to describe some general properties which
apply to a hypothetical filamentous organism, keeping in mind that every specific case
will somehow differ from this scheme. 

\begin{figure}[h]
\centerline{
\epsfxsize=8cm
\epsffile{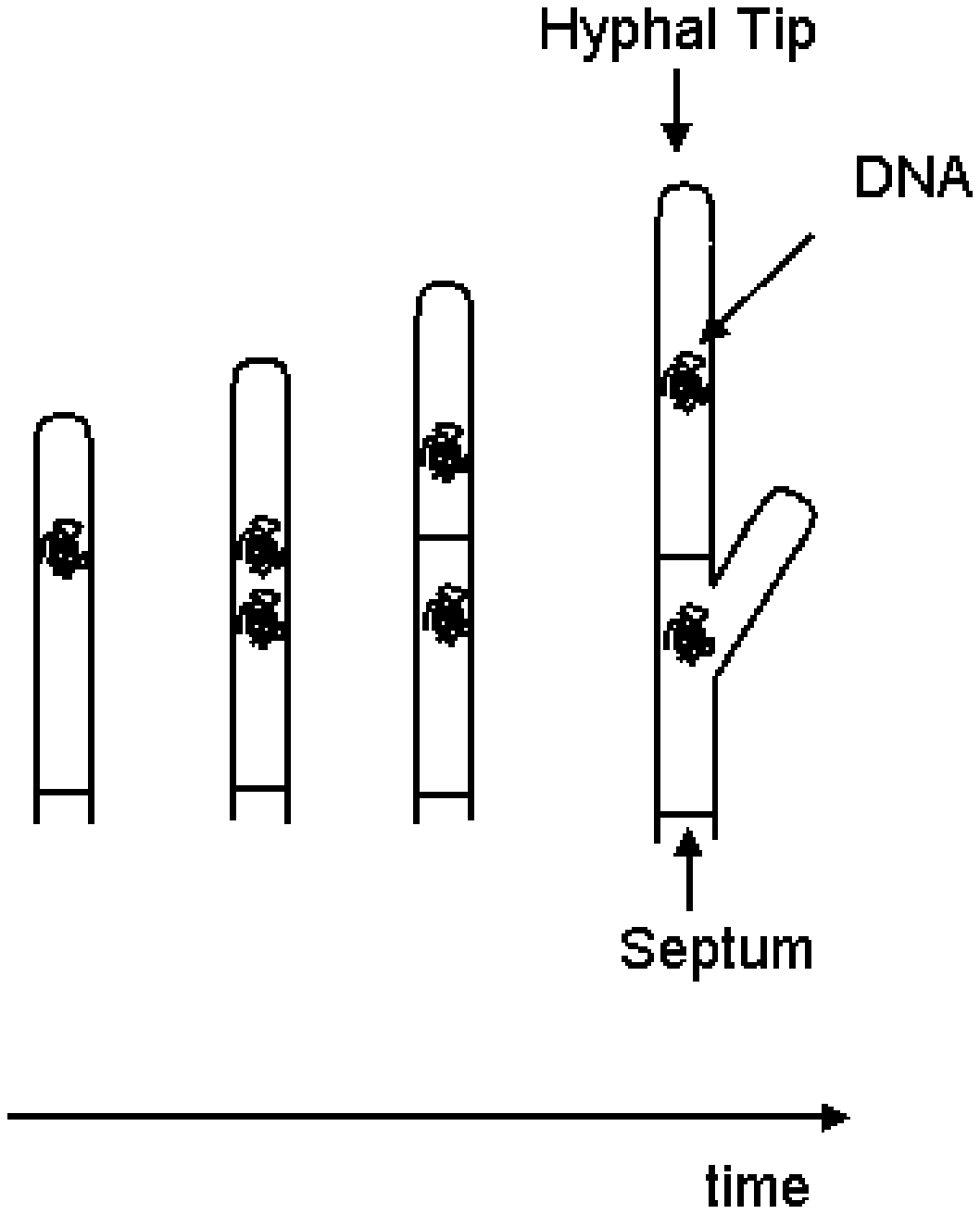}
}
\caption{\label{Fig:hyphal_growth}Filamentous organisms are organized in
hyphae, where single cells are separated by septa and growth is localized at
the hyphal tips. As apical cells grow, DNA is replicated, and eventually a septum
separates mother and daughter cells. Occasionally, subapical cells can undergo
a branching process.}

\end{figure}

A colony originates from a single spore, which germinates in favorable
conditions. Cell elongation
is localized to the tips, where new cell wall is continuously
assembled~\cite{Mig,Gooday}. In a typical cell division cycle, Fig. \ref{Fig:hyphal_growth}, the cell grows,
replicates its DNA, and the genetic material is segregated to
mother and daughter cells. Finally, once the mother cell has reached a critical size
(roughly, it has doubled its original size), 
septum formation separates mother and daughter cells ~\cite{Kretschmer81,Fiddy76}. 

This cell division process
gives rise to an apical  and a subapical cell. They differ greatly, since the apical cell
is actively growing, growth being localized again at the hyphal tip,
while the subapical cell does not grow, but supports tip growth
by producing material that is delivered to the apical cell.
Successive rounds of coordinated elongation
and septa formation produce a hypha formed by several different cells
separated by septa~\cite{Kretschmer81}. Notice the difference 
of this process from cell division occuring in cells that replicate by 
binary division: in a colony of E. coli every cell of the population can replicate. In
filamentous organisms the very mechanism of cell division 
implies differentiation for cells located in
different positions along an hypha.

The hypothetical organism we have described so far is
formed by a hypha growing at the tips.
Extension rate increases exponentially during the first hours,
synchronous with periodical rounds of DNA synthesis, and 
with the lengthening of
the supporting hypha~\cite{Kretschmer88}.
Tip growth rate reaches a maximum when growth rate equals the rate of transport to the tips
of material needed for cell wall construction. After this phase, tips growth cannot
match the production of cell wall material and elongation rate is
constant. However, the excess material   is not wasted, but used by
a second growing point branched from the hypha, Fig. \ref{Fig:hyphal_growth}.  
The very same growth properties
described for the original hypha apply to the new branch: cell growth
is localized at the tips, new tips can be formed by branching. 

Generally, in rich media branching
is highly favored, while it is inhibited in poor media.
However, during early growth, both the total number of branches  and
the total mycelial length  increase exponentially with the same
specific growth rate ($\mu$), so that
the colony expands exponentially. Their ratio is called the hyphal
growth unit, and it reaches a constant value after an oscillatory
transient ( a more quantitative analysis of this phase
of growth is given in Section~\ref{sec:Fungi-model}).

The relationship between branching,
growth and septa formation varies with the species and with the strains,
both in actinomycetes and fungi.
For example, {\it Streptomyces granaticolor} shows a pattern similar to fungi
where branches are localized near to septa 
\cite{Kretschmer81}, while in {\it Streptomyces coelicolor} branching occurs 
far from them \cite{Prosser91}. Moreover, in {\it Streptomyces coelicolor} apparently 
branching and septation can be completely separated, as mutants unable 
to septate are still capable of branching, and develop a vital mycelium
\cite{Mccormick94}.

\subsection{Colony morphology and kinetics}
\label{bio::kinetics}

\begin{figure}[h]
\centerline{
\epsfxsize=8cm
\epsffile{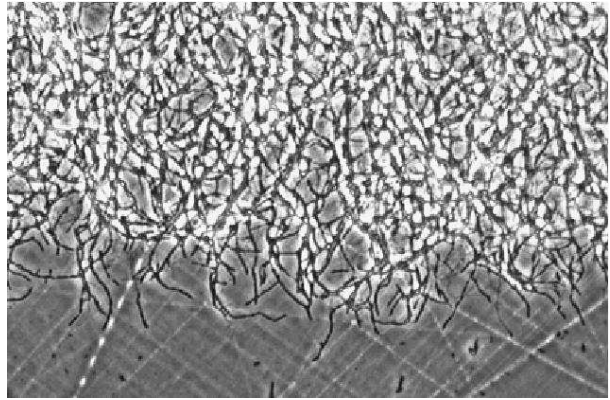}
}
\caption{Vegetative mycelium at the colony border of {\sl S. rutgersensis}: notice hyphal tips 
growing outwards.}
\label{foto:mycelio}
\end{figure}

After a few hours, a colony usually develops a circular shape,
and the colony radius elongates with a linear extension rate,
driven by the growing tips at the colony margins (see Fig.~\ref{foto:mycelio}).
More precisely, it is possible to define a ``peripheral growth zone'' 
with width $w$~\cite{Trinci71}, actually a ring of hyphae, that is actively involved in 
colony elongation. Both hyphal
tips, where the growth process occurs, and supporting cells, that
produce material needed for growth to be delivered to the tips, belong
to the peripheral growth zone.

An empirical equation describes
the relationship between colony radial rate ($dr/dt \equiv K_{\mathrm{r}}$,
at this stage, colony radius $r$ is an obvious choice to measure
colony growth), $w$ and the specific growth rate $\mu $~\cite{Trinci71} 

\begin{equation}
\frac{dr}{dt} \equiv K_{\mathrm{r}}=\mu_r w
\end{equation}

This empirical relationship is useful to understand the strategy adopted
by filamentous organisms to exploit resources~\cite{Prosser91}. Both in a rich and poor environment,
a colony grows to completely cover an
agar dish,
but the total biomass produced is different in the two media. When the environment
is poor in nutrients, branching is kept low, the number of tips is
low and therefore $w$ can be kept high. This way, even though $\mu $
is small, the colony moves with an high extension rate $K_{\mathrm{r}}$
through a region that cannot provide enough nutrients to maintain a
large number of cells. That is why, for example, {\it Streptomyces} can fill a whole
petri dish even growing in minimal media.
On the other hand, if the area is rich in nutrients,
branching is very high, to exploit completely the available resources.
The resulting $w$ is small, because many growing tips need to be
supported, but the extension rate $K_{\mathrm{r}}$ is 
kept high by a large $\mu$. A very similar strategy is adopted by fungi, but the
dimensions are different: $w$ in {\it Streptomyces} are typically one order
of magnitude smaller than in fungi.

Even though only cells located at the margins are responsible
for radial growth, growth occurs in the center of a colony as well.
In {\it Streptomyces granaticolor}, Kretschmer~\cite{Kretschmer81} reports
various cycles of DNA replication, branching and septation 
in subapical cells,
as certified by the smaller average size of subapical not branched
cells compared to apical cells. 
However, eventually, 
branching and  septation cease completely in the center of a colony.
Cells located in this area are the first to detect a lack of nutrients 
and the presence of a high concentration of catabolites \cite{robson87}. Not surprisingly some of them
differentiate into aerial mycelium, that will give rise to spores. These specialized cells
will be transported by physical or biological vectors to new 
regions where they will develop into new colonies.

A detailed description of aerial mycelium formation is beyond the scope of
this comment (see~\cite{Chater97}); however, 
it is important to stress that differentiation relies upon
intercellular signaling. For example, in {\it S. griseus}
a substance termed A-factor is involved in triggering
aerial mycelium production~\cite{Hara82}. Extracellular diffusion of the
inducers of aerial mycelium guarantees a 
global coupling of the cell population. 
Aerial mycelium growth is supported by other cells that undergo lysis,
thus providing nutrients for the aerial mycelium~\cite{Mendez85}. At this stage,
{\it Streptomyces} produce 
antibiotics to prevent other bacteria to have access
to this source of nutrients~\cite{Chater92}.
Usually aerial mycelium grows in a compact circular
pattern but different patterns can develop as well.
{\it S. rutgersensis} grown on minimal media forms concentric rings of
aerial mycelium (see Fig.~\ref{strepto-maxmin})~\cite{Bezzi}. Similar
but order of magnitude larger rings are also observed
in  {\sl Neurospora crassa} \cite{deutsch93}. Mathematical models have been
produced to explain the generation of these patterns: they will be presented in
Section ~\ref{sec:aerial}.

%%%%%%%%%%%%%%%%%%%%%%%%%%%%%%%%%%%%%%%%%%%%%%%%%%%%%%%%%%%%%%%%%

\section{Modeling fungal growth }
\label{sec:Fungi-model}  

\subsection{Macroscopic models}

One of the first empirical macroscopic model was proposed by Trinci~\cite{Trinci74}
to explain mycelial growth in its early stage.
Let us call $\psi(t)$ the number of branches and $\lambda(t)$ the total 
mycelial length at time $t$. During the early stages
of growth, the temporal evolution of
these quantities is well reproduced by the following equations:

\begin{eqnarray}
\psi(t)&=&a \exp (\mu t)
\label{trinci1}\\
\lambda(t)&=&c \exp (\mu t)
\label{trinci}
\end{eqnarray}
with $a$ and $c$ constants and $\mu$ the growth rate.
From Eqs.~(\ref{trinci1}-\ref{trinci}), we immediately get:
\[
\frac{\psi(t)}{\lambda(t)}=G
\]
that is, the ratio $G$ (called {\it hyphal growth unit}) is  constant over time.

These phenomenological relations are in agreement with growth patterns of
several filamentous fungi (see for example~\cite{Bull77}) and bacteria
(see for example, Figure 2 in ~\cite{Allan83}).

During the initial phase of growth, after a short lag period, also
the radius of the colony $r_c$ grows exponentially.
(see Ref.~\cite{Kotov90} for a model of this phase
of growth). Then, as described in Section~\ref{sec:biology}, linear radial growth sets up
while hyphae within the interior of the colony do 
not grow further~\cite{Trinci74}. 
During this phase, there is a quadratic increase in biomass.
In fact,  
\[
r_c \propto  t
\]
(with $t$ the time from the beginning of linear growth phase)
and assuming  biomass $M$ uniformly distributed throughout the colony, 
(i.e. the total biomass  proportional to the colony surface).
\[
M \propto r^2 \Rightarrow M \propto t^2
\]
That is, total biomass grows quadratically in time.

This simple model found many experimental confirmations,
but it does not supply any information on the local 
distribution of the biomass, and in particular on the different
behaviors between center and border of the colony.
Using a reaction-diffusion model Ferret {\it et al.}~\cite{Ferret99} have predicted
the evolution of the biomass and tips density  under different growing conditions.
In presence of unlimited nutrients and no spatial constraints, 
Eqs.~(\ref{trinci}-\ref{trinci1}) are recovered, and the biomass increases exponentially 
with time. 
However, as mentioned above,
growth stops  in the center of the colony: in the model
the same happens when the biomass 
density $m(x,t)$ reaches a maximum value, $m_{max}$, while at the colony border
the growth continues freely. 
These differential growing velocities are modeled by introducing a 
{\it collision probability} $P_c$ as a function of local 
biomass density $m(x,t)$. 
In the the central part of the colony, the local density
quickly reaches its maximum value, $m(x,t)=m_{max}$, 
there is no free space available 
for growth, collisions are unavoidable and the free space (proportional to 
$1-P_c$) is zero: the mycelium cannot grow further.
On the other hand, at the colony border, no collisions happen, $P_c=0$,
and the hyphae can grow freely outward.
Assuming a linear model for $P(m(x,t))$ (see Fig.~2 of the cited paper),
that fulfills these two limiting cases, the model is able to predict the
temporal behavior of total biomass and the colony border evolution
(i.e. internal and external radius, defined as the radius of the not-growing
core, and the radius of the whole colony, respectively).
The model has been validated by comparing its predictions to the behaviors of 
two different fungi: {\it Gibberella fujikuroi} and {\it Aspergillus
oryzae}.

These macroscopic models are suitable for 
modeling the growth of the whole colony, but they are not able 
to grab all the complex fine structure of mycelial growth.
In fact, in addiction to tip extension and branching,
filamentous  growth is characterized by septation and 
internal transport mechanisms ({\it translocation}): these microscopic
features will be taken into account in the next class of models.

\subsection{Microscopic models}

\paragraph{Septation and branching}
 
Once the nuclear material is replicated, mycelial microorganisms divide a
hyphal compartment in
two parts by septation, Fig.~\ref{Fig:hyphal_growth}.
 Branching  usually occurs after septation, with
a fixed time delay (see for example the case of {\it 
Geotrichum candidum} in ~\cite{Fiddy76}).
A model that includes branching and septation has been introduced
by Yang {\it et al.}~\cite{Yang91}
The model includes a deterministic part, that mirrors tip growth and 
septation, and a stochastic part to keep track of new branch 
initiations. 
Septation is triggered when the amount of nuclear material 
has been doubled and segregated.
Due to the random behavior of branching, new branches are supposed
to be generated at random around their respective septa according
to a normally distributed probability with a fixed variance
$\sigma^2_{L_b}$. In the limiting case, $\sigma^2_{L_b}=0$, 
new branches grow just behind septa. 
Tip growth and new branching directions are calculated according to
a random model described in~\cite{Yang92}. 
In short, 
experimental observations in 2-D colonies reveal that tip growth angles
 $\Delta \gamma_g$ and branching outgrowth angles $\Delta \gamma_b$ are
normally distributed around zero.
In the model, that represents growth in a 3-D environment.
The orientation of the primary hyphal segment  with respect to the straight
direction is  obtained  sampling the experimental $\Delta \gamma_g$ distribution.
In the plane perpendicular to the growth axis, the direction is fixed by an additional random
angle $\theta$ assumed to be uniformly distributed between $[0, 2\pi]$.
A  simile procedure is used for determining the branching angle $\Delta \gamma_b$
in the plane
perpendicular to the hyphal growth axis (see Fig.~\ref{Yang:angle}).

Simulation results are in good agreement with experimental data for
the fungus {\it Geotrichum candidum} (taken from Ref.~\cite{Fiddy76})
and for the filamentous bacterium {\it Streptomyces tendae}.
 Both germ tube growth and
temporal behavior of the number of tips and septa can be reproduced.
In addition, the morphology of the growing mycelium
appears similar to that observed in experiments (see Fig.~10 in~\cite{Fiddy76}).

\paragraph{Translocation}

Another peculiar aspect of 
the internal kinetics of a growing mycelium is the capability
to distribute nutrients through the  mycelial 
network (translocation), both
passively (due to diffusion through the hyphae) and actively
(due to active and hence energy-dependent metabolism)~\cite{Olsson98,Persson00}.  
This way, fungi  can absorbe nutrients 
in some parts of the mycelium and transport them through the network,
particularly where food is  scarce.
This mechanism can play a crucial role for colonies growing
in highly heterogeneous environments, such as most of the natural ones.

 Among those that have been recently proposed~\cite{Davidson98,Davidson00}, we will
describe here a recent model introduced by Boswell {\it et al.}~\cite{Boswell02}
to study the fungus {\it Rhizocotonia soloni}. 
The model includes five components (densities):
active and inactive hyphal densities, tip density, internal and external
substrate concentrations. 
Inactive mycelia, usually the center of the colony, do not contribute to translocation, while 
translocation occurs in active hyphae.
Internal passive translocation is modeled as a diffusion-like
term for internal substrate, $s$, that depends 
on mycelial concentration $m_a$,
i.e. in presence of a dense mycelial network the 
diffusion process is much faster
than in a sparse one. Active translocation is modeled
by assuming that active mechanisms are due to
the continuous substrate demand by growing tips, i.e. 
the flux follows the gradient of tips density $p_t$.
So, active translocation moves substrate
from  low-density tip areas to high density ones (so called 
{\it tip-driven} diffusion) through the mycelial network 
(i.e. it also depends on $m_a$). 
In mathematical terms:
\begin{equation}
\begin{array}{r@{\quad:\quad}l}
%\frac{\partial}{\partial x} & \mbox{Passive translocation}\\
%\frac{\partial}{\partial x} & \mbox{Active translocation}
\mbox{Passive translocation}&\frac{\partial}{\partial x}\left( -D_i m_a \frac{\partial s}{\partial x}\right) \\
\mbox{Active translocation}&\frac{\partial}{\partial x}\left( -D_a m_a s\frac{\partial p_t}{\partial x}\right) 
\end{array} 
\end{equation}

The parameters are obtained from experiments.
Using this model the authors explore the interplay between active and passive translocation in 
a growing colony in two kinds of growth media, homogeneous (nutrients are homogeneously spread on the dish)
\footnote{The effects of different, but homogeneous, growth media
have also been  studied
in {\it Aspergillus Oryzae} colonies~\cite{Lopez03,Matsuura92,Matsuura93},
and  will be discussed further in the case of {\it Streptomyces} colonies in Section~\ref{sec:aerial}.}
  and heterogeneous
(substrates are patchy distributed).
Simulations suggest that diffusive translocation is used 
for random exploration while tip-driven translocation is used for optimizing the distribution
of available resources.
This view is coherent with the idea that optimal search combines a random
process with some more complex, 
but usually more demanding, strategies of resource exploitation.

\paragraph{Single hypha models}

Going down in the length scale, we find
single hypha models.
The average kinetics implemented in the previous models  
do not describe properly the evolution 
of individual hyphae.
 Indeed,  
due to the complexity of the hyphal structure and the
different growth rates of
individuals branches within an hypha (see \cite{Trinci74}),
  stochasticity has to be taken into account. 

\begin{figure}[h]
\centerline{\psfig{figure=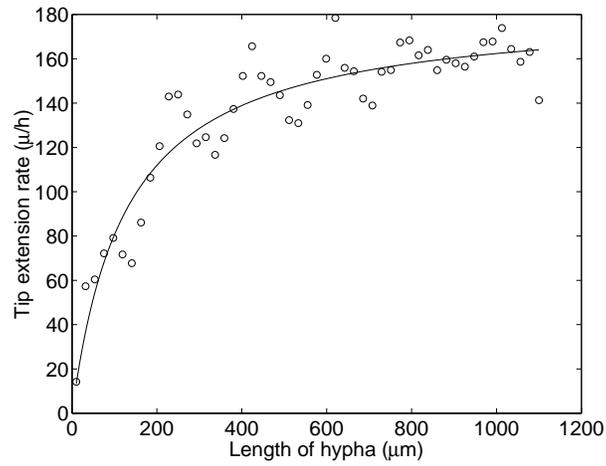,width=8cm}}

\caption{Tip extension rate for a single hypha, see Eq.~(\ref{qs}).
 Experimental
points are spread around the solid line, with temporary decreases corrisponding to 
branching point (adapted from Fig.~2 in Ref.~\cite{Christiansen98}).
}
\label{fig:qs}
\end{figure}

\begin{figure}[h]
\centerline{\psfig{figure=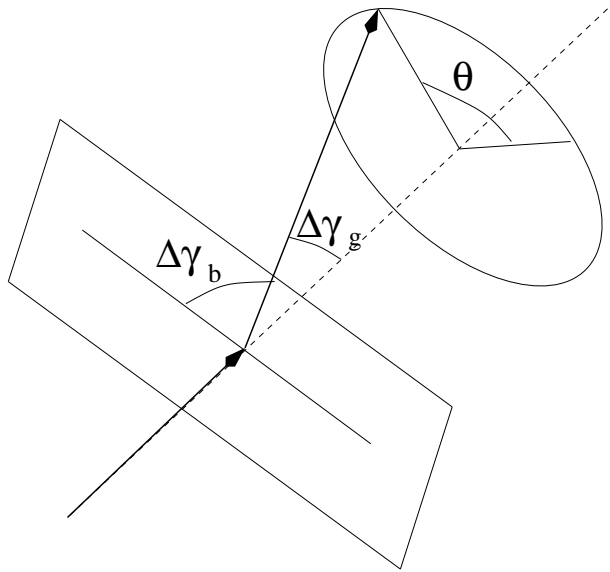,width=8cm}}

\caption{Schematic representation of growth direction angle and branching direction angle.
Adaptation from~\cite{Yang92}.
}
\label{Yang:angle}
\end{figure}

 Christiansen {\it et al.}~\cite{Christiansen98},
studied the growth of a single hyphal element of {\it Aspergillus oryzae},
combining image analysis
with a model that makes use of Monte Carlo simulations.
They consider the growth starting from a single spore; 
at each time step  ($\delta \tau$ ), the probability of
branching is: $exp(q_{bran}\delta \tau)-1$.
The branching rate $q_{bran}$ is computed from a phenomenological model
of tip kinetics $q_{tip}$\footnote{
A detailed model of
the elongation process of the growing tip
can be found in Ref.~\cite{Regalado99}. 
}, and from observed distribution of the number
of new branches as function of hyphal length $n_l$.
From experimental observations, tip velocity as function of
hyphal length $l$ can be fitted by the following function (see Fig.~\ref{fig:qs}): 
\begin{equation}
q_{tip}(l)=k_{tip}\frac{l}{l+K_s}
\label{qs}
\end{equation}
where $k_{tip}$ is the maximum tip
extension rate, $K_s$ is an empirical saturation constant,
that depends on environmental conditions (glucose concentrations),
and varies among different types of branches ($K_s$ is usually larger for the
germtube than for secondary hyphae).

In the model $K_s$ is calculated from:
\[
K_s=exp(\beta p + \alpha)
\]
with $p$ a normally distributed random number, and the parameters
$\alpha$ and $\beta$ fitted from experimental values of $K_s$.
Eq.~(\ref{qs})  is a good approximation of the average behavior 
of tip growth rate $q_{tip}$; but detailed experimental measurements show 
that a decrease in tip extension rate occurs in correspondence to
the formation of  new apical branches (see Fig.~\ref{fig:qs} adapted from Fig. 2. 
in Ref.~\cite{Christiansen98}, but also the discussion in Ref.~\cite{Hutchinson80}
where no correlations between tip extension rates and branching
are found). 
Following Ref.~\cite{Christiansen98}, we will neglect these fluctuations in the current
analysis.
After the hyphal  length has reached a certain 
threshold $l_{min}$ (that depends on the individual hypha, 
and differs greatly between 
primary, germtube, and secondary hyphae), the number of new branches
 $n_l$ formed on a hyphae 
grows linearly as function of length:
\begin{equation}
n_l= \left\{ \begin{array}{r@{\quad:\quad}l}
0 & l<l_{min}\\
k_b (l-l_{min}) & l>l_{min}
\end{array} \right.
\label{branchs}
\end{equation}
where $k_b$ is a branching constant (actually, it assumes two different
values for the germtube and for secondary hyphae).
From Eqs.~(\ref{qs}) and ~(\ref{branchs}), the branching intensity (i.e. 
the number of branches per time unity) can be derived:
\begin{equation}
q_{bran}=\left\{ \begin{array}{r@{\quad:\quad}l}
0 & l<l_{min}\\
k_b\frac{l k_{tip}}{K_S+l} & l>l_{min}
\end{array} \right.
\label{qbran}
\end{equation}
Using these values of $q_{bran}$ we can simulate 
the growth of a single hyphal element using a standard Monte Carlo
technique. In a standard simulation, a single spore has been considered, 
 the tip grows  with a rate $q_{tip}$ and at each time step $\delta \tau$ a new 
 branch is formed with a probability $exp(q_{bran}\delta \tau)-1$.
Although hyphae grow only at the tips, with a saturating velocity,
the total length of mycelium increases exponentially due to 
the exponential increase in the number of branches (after the 
minimal length $l_{min}$ has been reached).
The simulated growth kinetics agree with the observed experimental
patterns for the total length and the number of tips
for a single hypha.

A detailed study of single hypha growth has been  
presented as well by Hutchinson {\it et al.}~\cite{Hutchinson80}.
They record the temporal evolution of 
primary hyphae and first and second order branches  
in six colonies a few hours after germination.
Using the experimentally observed distributions, they 
proposed a  model of mycelial growth:
hyphae perform a straight-line growth for a length sampled
from an experimental fitted distribution.
At this point, branching can occur, and  
the direction for the new branch is obtained by
sampling from experimental distributions.
This model is capable of reproducing the morphology
of the early stages of {\it Mucor hiemalis} growth,
and represents one of the first model where
a single hypha dynamics (microscopic) is used for generating
macroscopic growth patterns (morphology of the mycelia).

From the single hypha level, we could go down one level in the analysis to
the genetic and molecular basis 
of filamentous growth 
(i.e. to the genetic networks underlying these processes).
However, although progresses have been recently made in elucidating the molecular
basis of mycelial growth and development (for example, see \cite{Chater98} for
a review on {\it Streptomyces}), no mathematical link between
physiology of cell behavior and molecular interactions have been successfully
produced so far.

\subsection{Cellular Automata}

\label{CA}
Cellular Automata (CA) are
fully discrete dynamical systems, i.e.  their 
dynamical variables are defined at
the nodes of a lattice (spatial discretization), with values taken from 
a finite set (state discretization) and evolve in discrete time steps
(temporal discretization).
 They were
introduced more than half  century ago by John Von
Neumann's~\cite{VonNeumann66}. As the choice of the name suggests,
the original idea is 
that the complex behavior characterizing many biological 
systems should originate
as a collective effect from simple local interaction rules (cell-interactions). 
Since then, CA have been applied to a vast range of biological and
physical problems (see for example the
reviews~\cite{Chopardbook,Bezzi01,Bagnoli98}
and the ``classical'' works \cite{VonNeumann66,Gardner71,Wolfram86})
\footnote{Another class of fully discrete models that emphasize 
local interactions as opposed to global ones are  Lindenmayer Systems
(L-systems)~\cite{Lindenmayer68,Prusinkiewicz89}. L-systems are basically a set of string re-writing rules
applied recursively. They are closely related to 
cellular automata, but usually they lack  an 
explicit spatial extension. These rules can mirror growth process,
and they are   generally used for generation of fractals and modeling of plants.
An applications to fungal growth, {\it Mucor hiemalis}, was proposed
by Soddel {\it et al.}~\cite{Soddel94}. We will not discuss these models here, but
we refer the readers to the cited paper and references therein.}

A CA model for fungal growth has been proposed by Lazlo \&
Silman~\cite{Lazlo93}. They consider a two dimensional 
square grid (that mirrors the dish) where each site has
$8$ neighbor sites (Moore neighborhood). Each site can be either occupied by
a mycelium  or empty.
A probabilistic rule is set for updating
the status of each site of the lattice simultaneously.
 In its simplest version the  rule is the following:
each empty site will become occupied with a probability $p_a$ dependent
on the number of neighbor occupied  sites $n_d$:
\begin{itemize}
\item $n_d = 1$ (one neighbor site occupied) $\Rightarrow p_a=0.12$
\item $n_d = 2$  $\Rightarrow p_a=0.25$
\item $n_d = 3$  $\Rightarrow p_a=0.50$
\item $n_d \ge 4$ (overcrowded) $\Rightarrow p_a=0$
\end{itemize}

Even using only this simple rule, interesting patters are generated, and
 the temporal behaviors of biomass and radius of the colony are 
 well reproduced.  This basic model has been extended 
 considering three different states of differentiation: vegetative
 mycelium, differentiating mycelia (an intermediary state) and
 spore formation, and setting the corresponding probability
 of transition according to local occupancy rules. 
Still keeping the model simple (one single byte  is enough to represent
the state of a site) the automaton is able 
to produce patterns for spore and mycelia, reminiscent of the 
experimental growth patterns.

One of the advantages of CA, is that they can be used
 in testing hypotheses about  the interactions among cells: 
some schematization (rules) of microscopic dynamics can
be easily implemented as cellular automata; then we can
observe the macroscopic patterns generated and compare it
with experimental patterns. 
This class of models is particularly useful when the macroscopic 
patterns observed do not
depend on the fine details of microscopic interactions,
but they emerge as a collective behavior of many 
cells (the site of the automaton).
In this framework, cellular automata, despite their simplicity, 
turn out to be a valuable tools for the studies
of emergent cooperative effects.

\section{Aerial mycelium pattern formation}
\label{sec:aerial}

As described in Section~\ref{bio::kinetics}, aerial mycelium  can form 
concentric ring patterns in  
{\it S. rutgersensis} colonies (see Fig.~\ref{strepto-maxmin}).
Ring patterns are found also in other 
bacterial and fungi colonies~\cite{Adler66}. 
Many models can produce such patterns~\cite{mats98};
in the case of bacterial colonies, it was suggested that
the interplay of front propagation and Turing instability
leads to concentric rings and spot patterns ~\cite{Tsimiring95}.
In the more specific case of filamentous bacteria, 
a different approach based on competition for resources has been
proposed by Bezzi {et al.}~\cite{Bezzi}. 
The model is based on the idea that 
growing hyphae  have the biological objective
to find nutrients to give rise to spores (aerial mycelia). As a result,
we expect a strong competition to arise on minimal media 
for the energetic resources between neighbor substrate mycelia,
whereas in
maximal media, where there are sufficient nutrients, the
competition is weaker.

If cells are connected mainly along the radial
direction (hyphae start growing mainly radially
from the initial spore), then competition will be stronger
along this direction than along the tangential one.
In other words, in the growing edge of the colony,
the competition is not isotropic
but, following the vegetative mycelium morphology, it will be stronger
among cells belonging to
neighboring circumferences (radial direction)
than among cells belonging to the same (tangential direction). 

\begin{figure}[f]
\centerline{
\epsfxsize=8cm
\epsffile{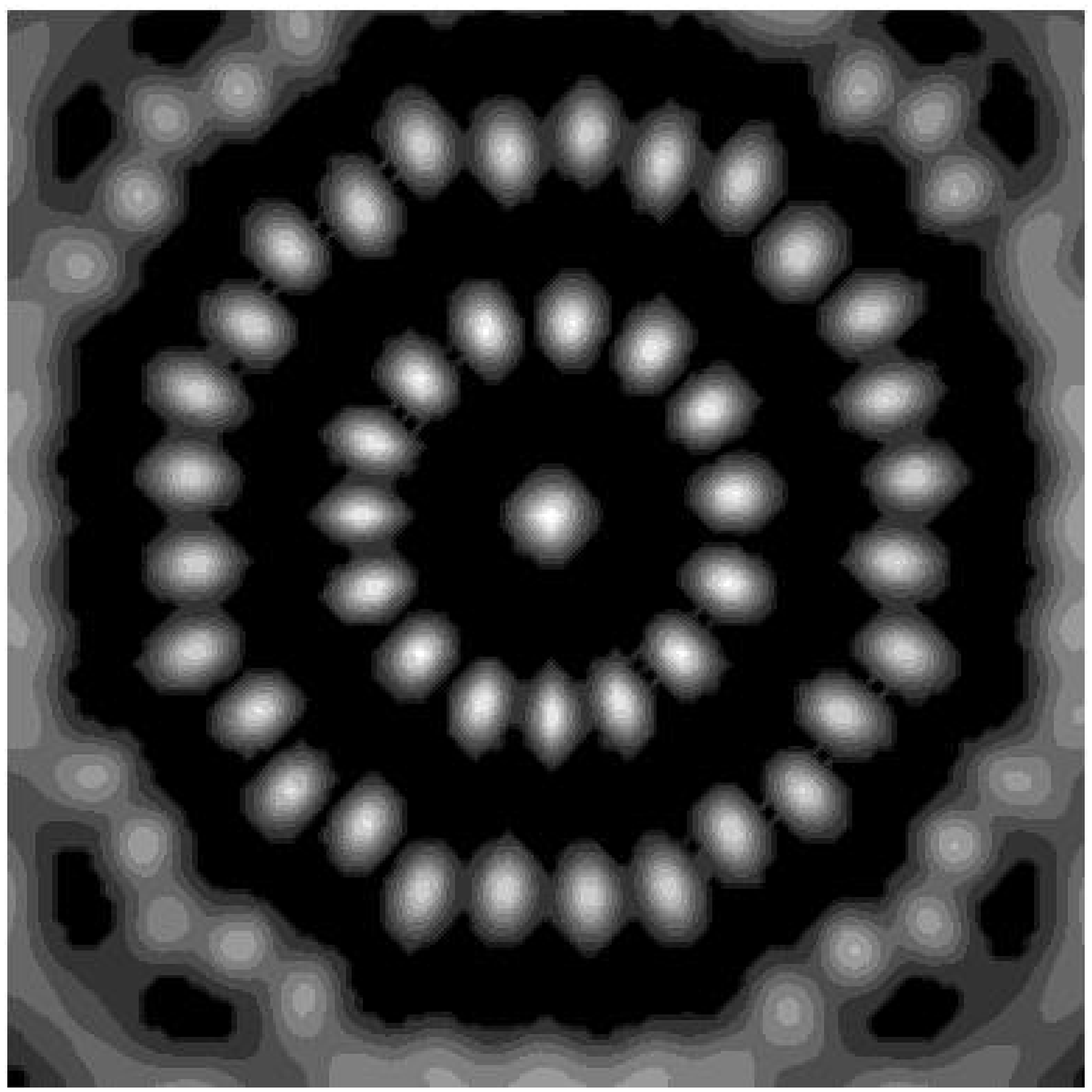}
\hspace{.5cm}
\epsfxsize=8cm
\epsffile{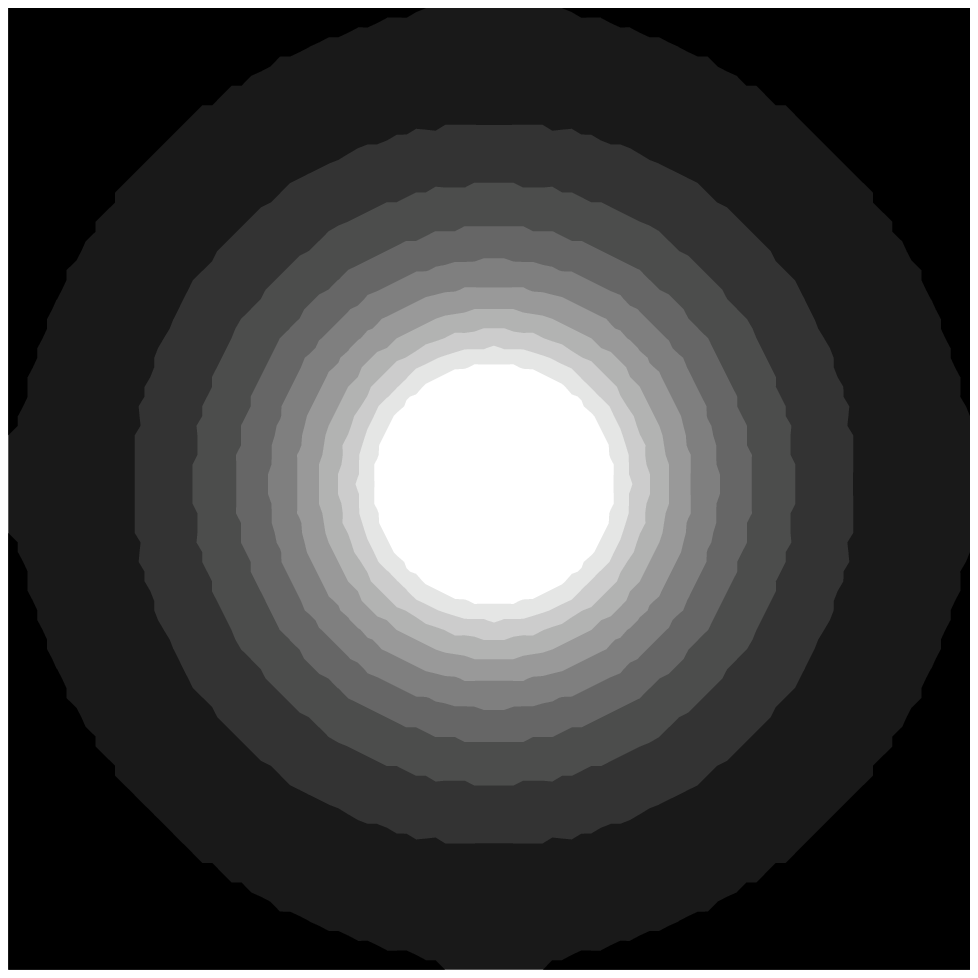}
}
\caption{Simulations of aerial pattern formation of {\sl S. rutgersensis}. Left, growth on minimal media,
right, growth on maximal media. From ~\cite{Bezzi}}
\label{Bezzi-maxmin}
\end{figure}

\begin{figure}[h]
\centerline{
\epsfxsize=8cm
\epsffile{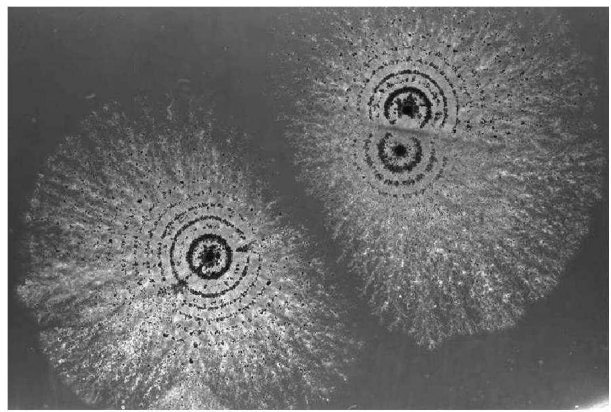}
}
\caption{Colonies of {\it S. rutgersensis} develop interference pattern when grown in proximity 
}
\label{strepto-int}
\end{figure}

\begin{figure}[h]
\centerline{\hbox{\psfig{figure=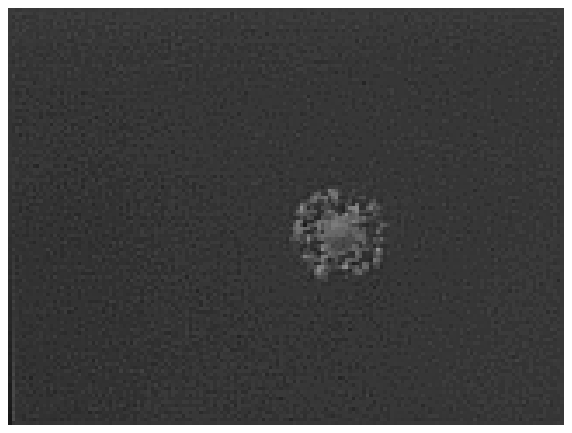,width=6cm}
\psfig{figure=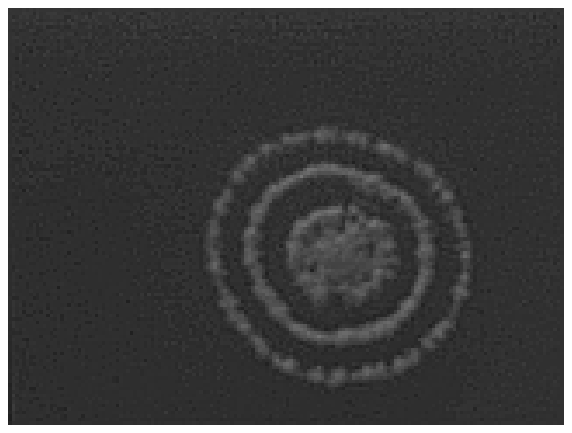,width=6cm}
\psfig{figure=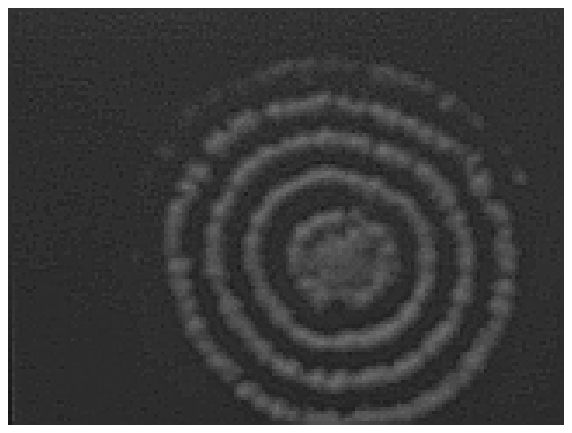,width=6cm}}}
\vspace{.5cm} 

\centerline{\hbox{\psfig{figure=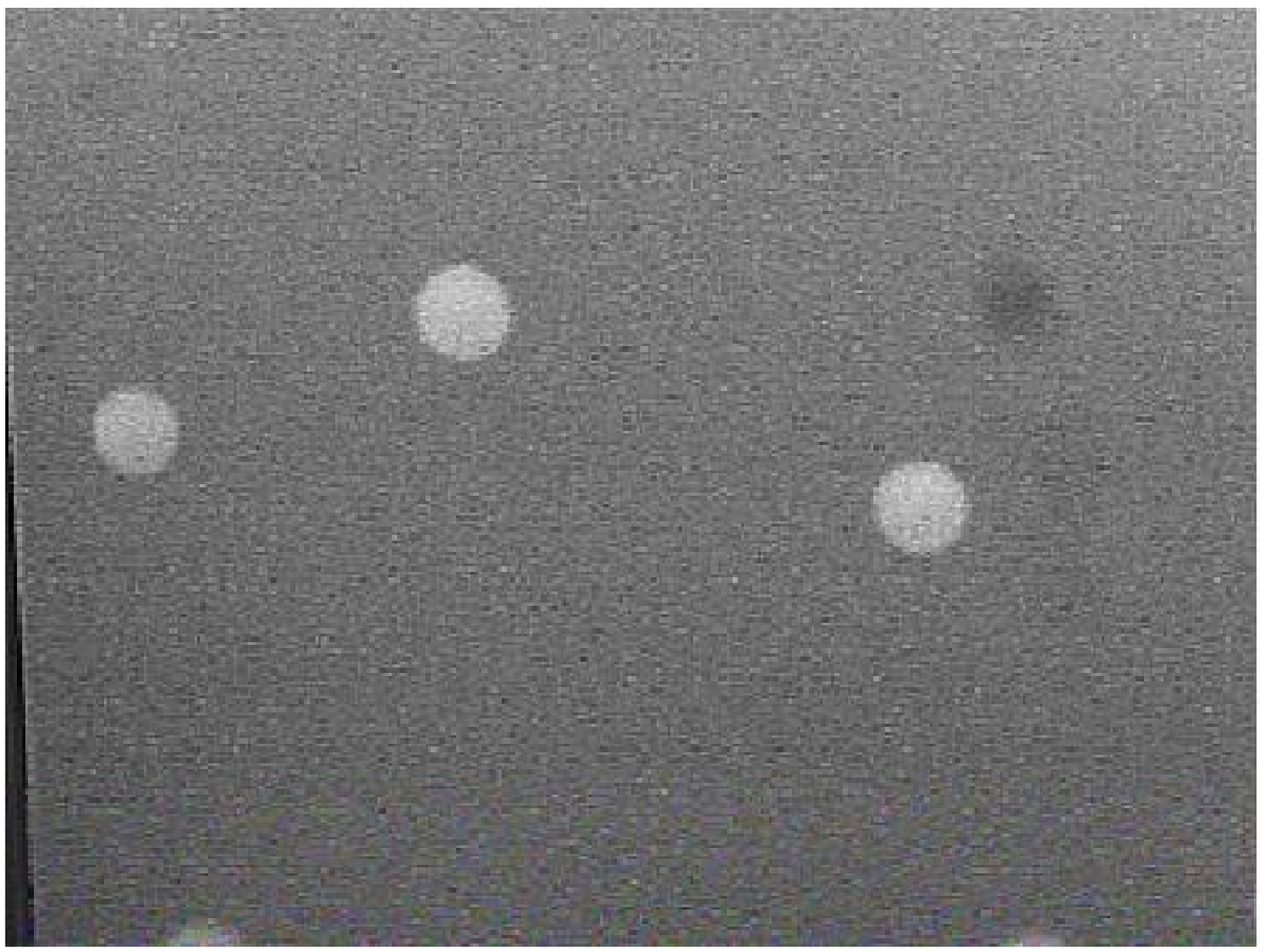,width=7cm}
\psfig{figure=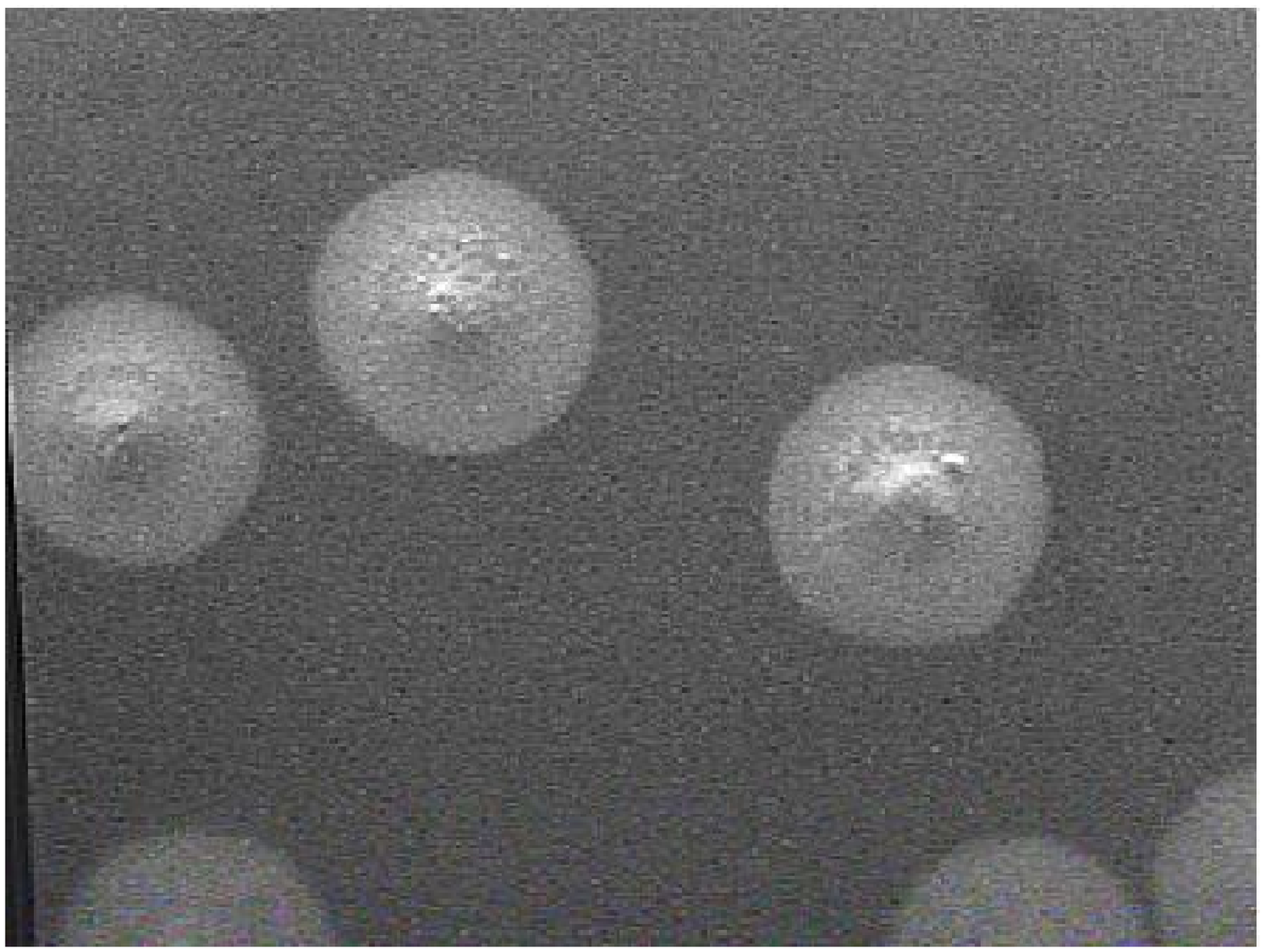,width=7cm}}}

\caption{{\it S. rutgersensis} growth in minimal (top) and maximal
 culture media (bottom, $3$ colonies are clearly visible).
 Aerial micelium: white spots; vegetative micelium not visible.}
\label{strepto-maxmin}
\end{figure}

\begin{figure}[h]
\centerline{\hbox{
\psfig{figure=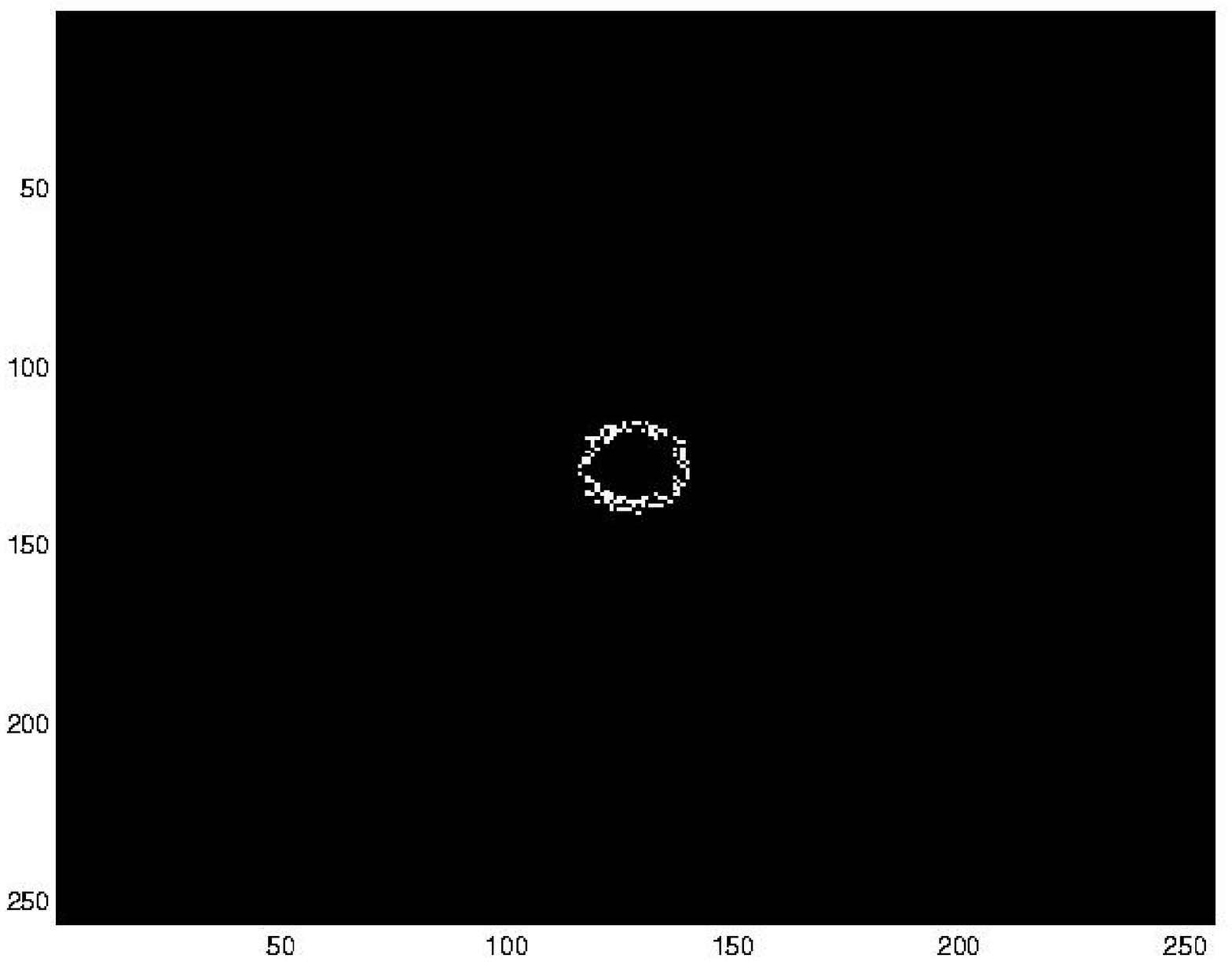,height=6cm,width=6cm}
\psfig{figure=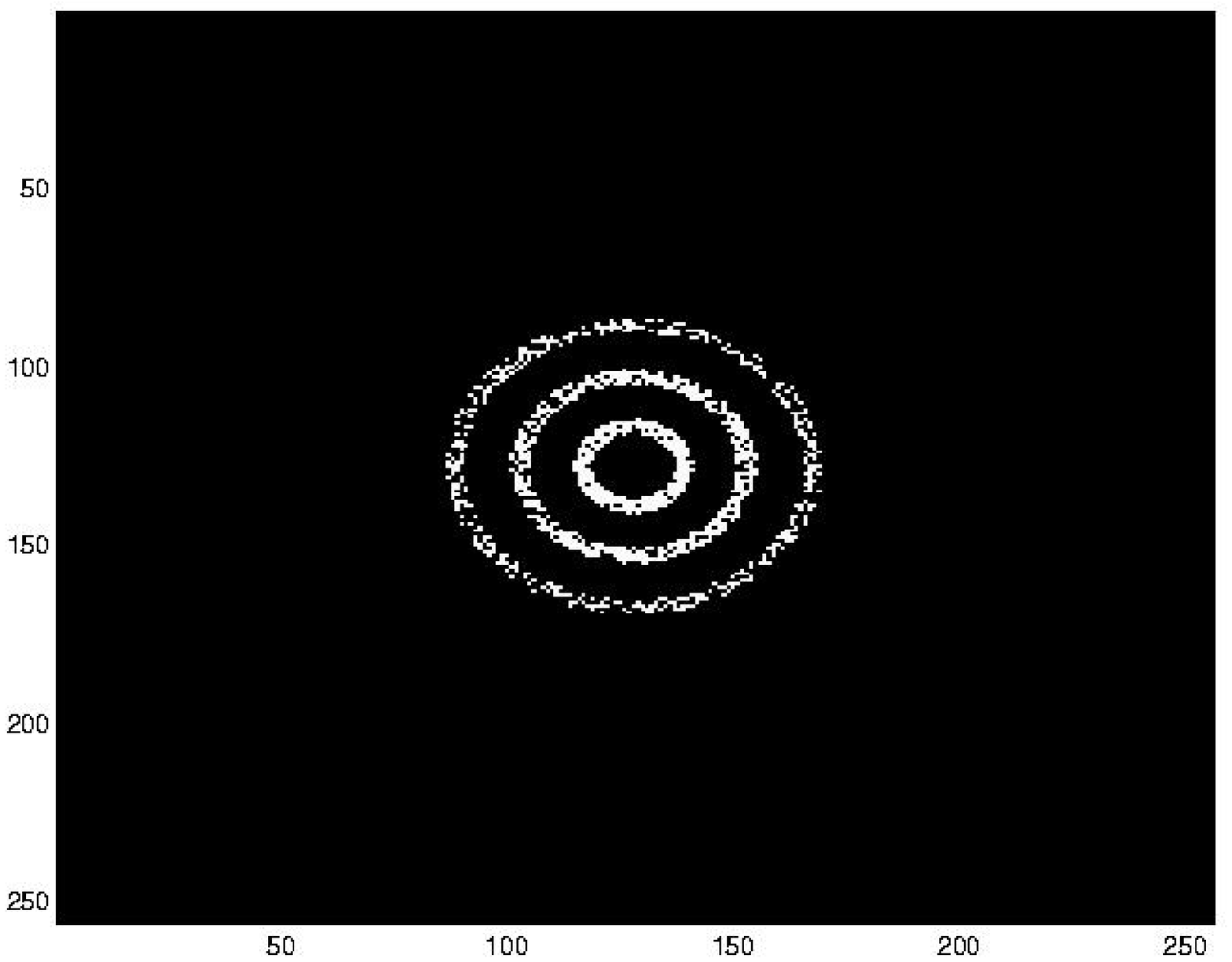,height=6cm,width=6cm}
\psfig{figure=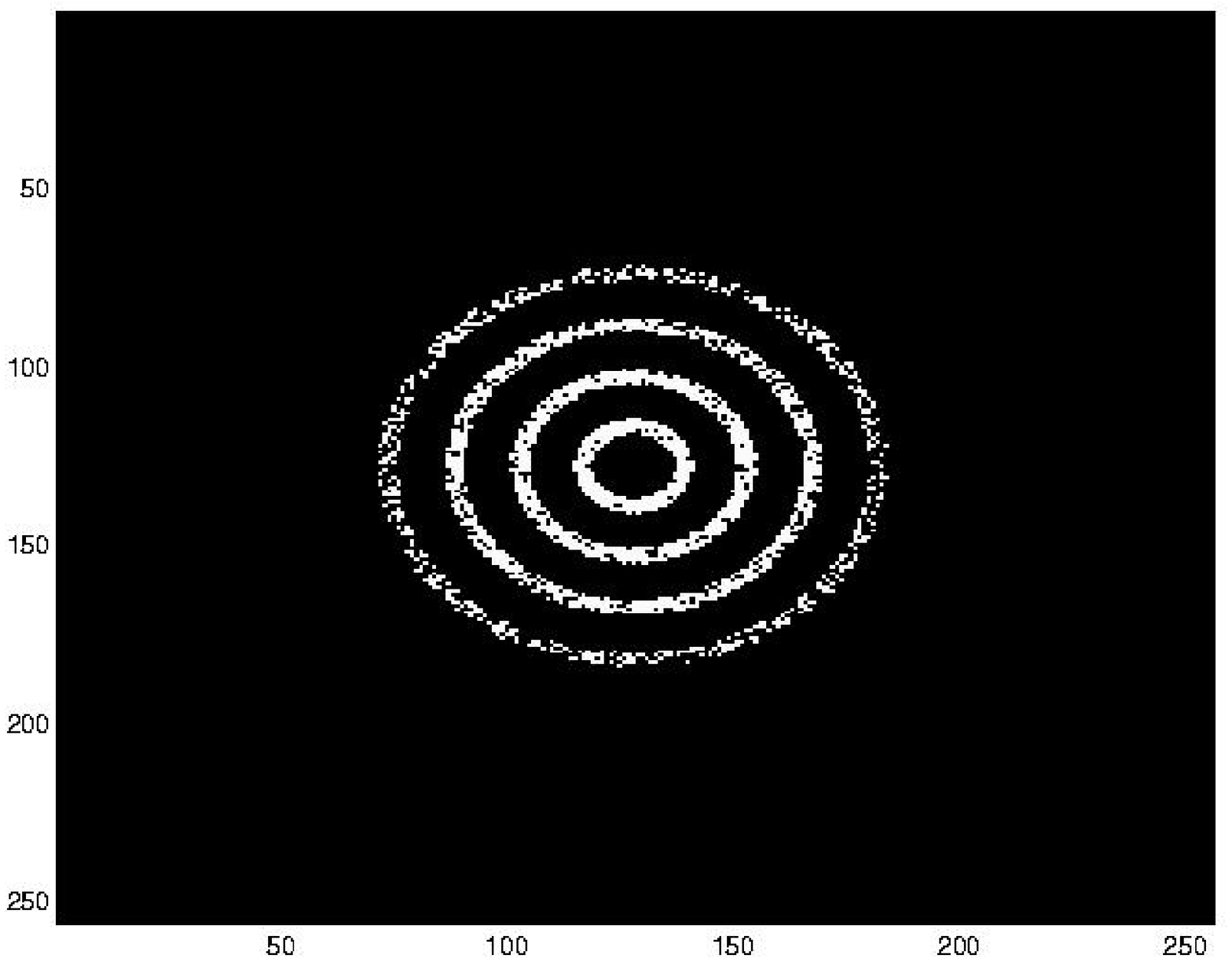,height=6cm,width=6cm}}}
\vspace{.5cm}

\centerline{\hbox{\psfig{figure=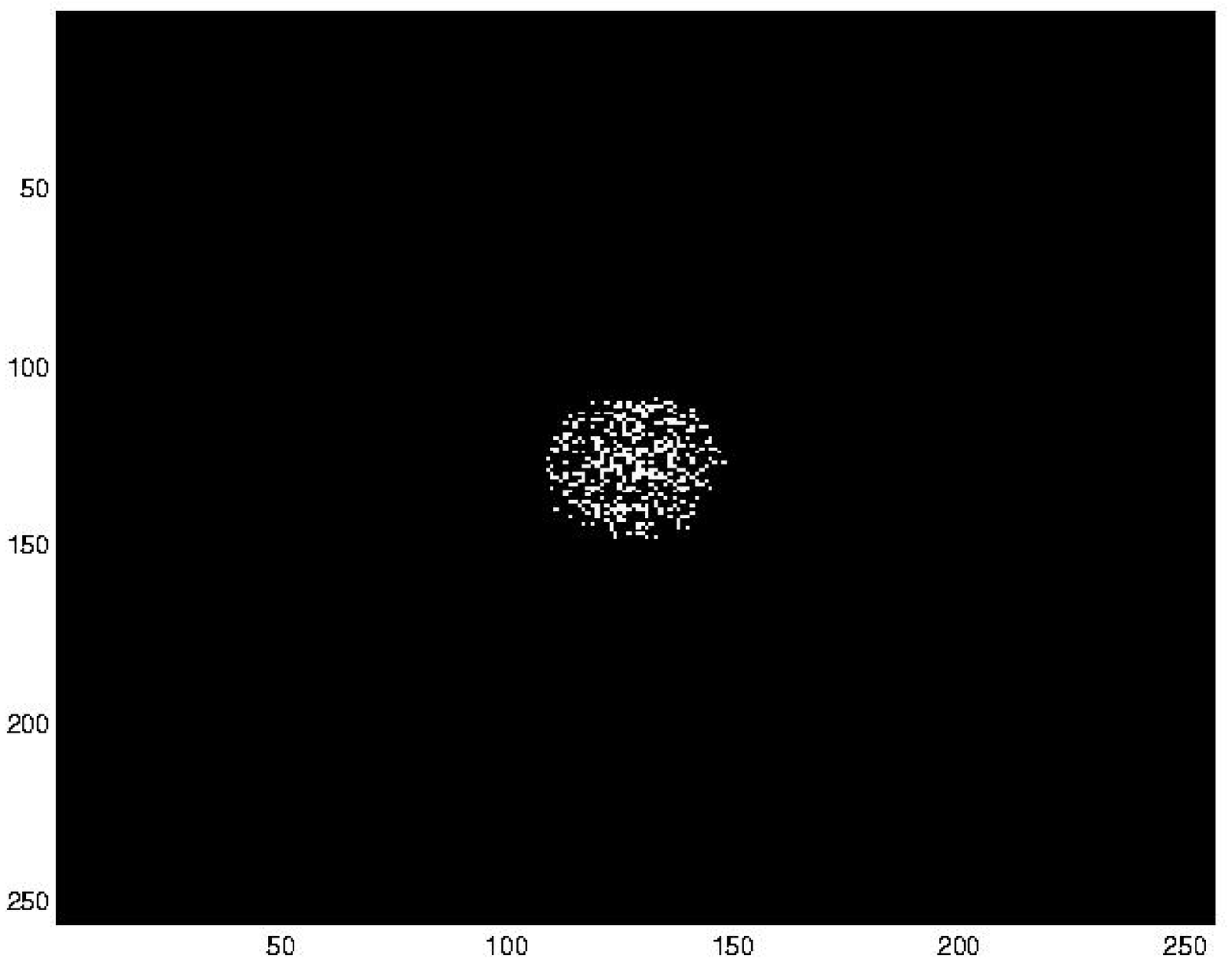,height=6cm,width=6cm}
\psfig{figure=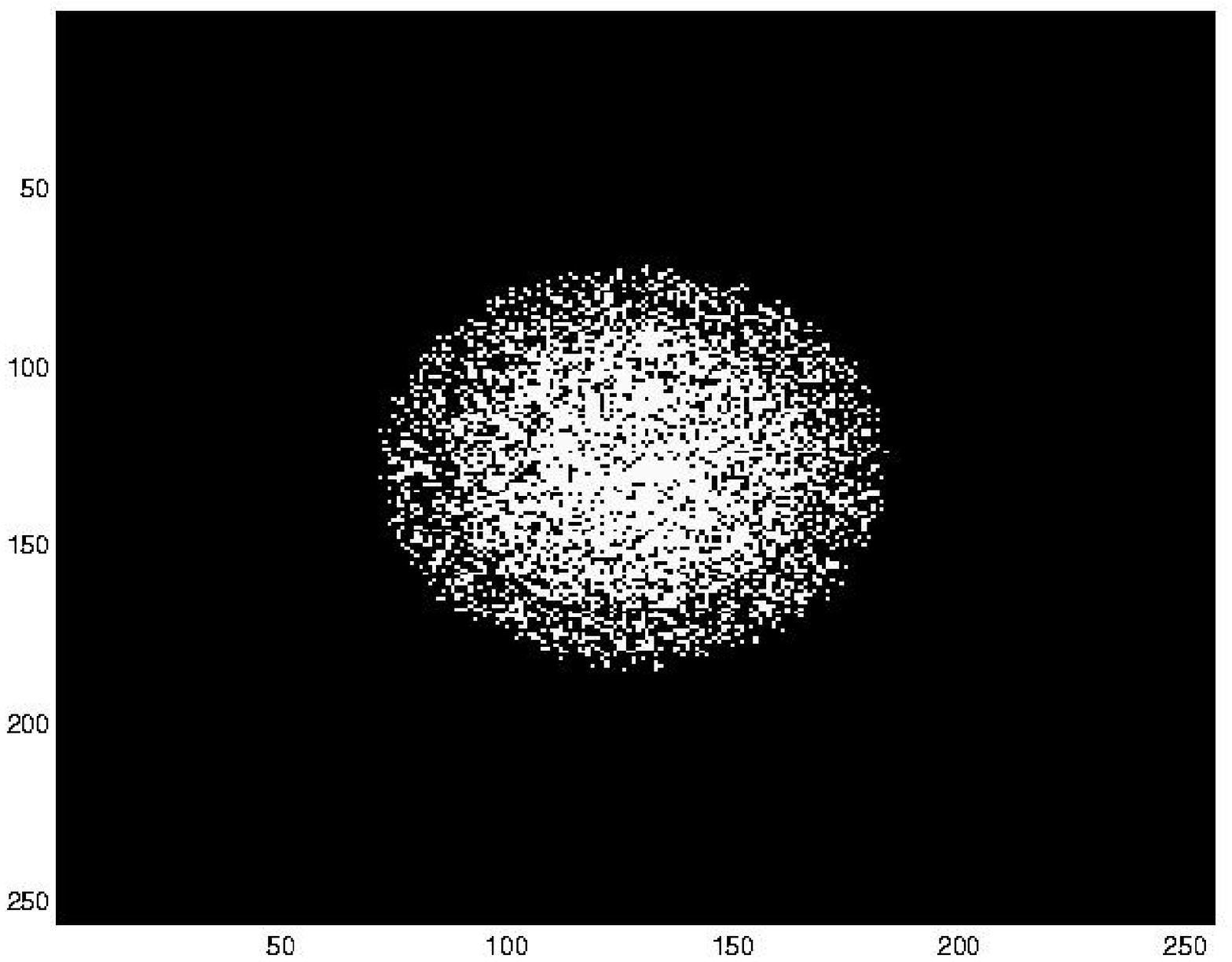,height=6cm,width=6cm}}}

\caption{(Top) Temporal evolution in minimal culture media
 Bottom:  Temporal evolution in maximal culture media
The model is based on a coupled-map lattice (a dynamical system discrete
in time and space), with four ''populations'': concentration of vegetative
mycelium and tips, concentration of food in the dish, concentration 
of food absorbed. These populations evolve in two different timescales:
a fast one, that includes tip growth, diffusion and translocation;
and a slower timescale for micelium duplication.
The combined signal of food scarcity and high absorbed food triggers
stochastic aerial micelium production (plotted in the figures).} 
\label{Bezzi-cm}
\end{figure}

In presence of high concentrations of nutrients, the competition is rather 
weak  and the final distribution of aerial mycelium
has the shape of a
gaussian-like distribution centered on the initial spore
(similar to the one shown in Fig.~\ref{Bezzi-maxmin}, right). 
If the colony grows in a nutrient-limited environment, 
hyphae that are neighbor along the radial direction compete 
strongly for nutrients,
 aerial mycelia development is inhibited,
and concentric ring patterns appear
(see Fig.~\ref{Bezzi-maxmin}, left).
Despite its simplicity (one differential equation) the model
is able to capture many peculiar features of {\it Streptomyces rutgersensis}
growth. However, since  the model does not include explicitly 
the growth of vegetative
mycelium, some important aspects, like the ``interference'' 
patterns among neighboring
colonies (Fig.~\ref{strepto-int}), are not reproduced.
Clearly, a more detailed view is necessary; a coupled maps model including 
vegetative mycelia and tip growth is under development~\cite{Cmaps} whose
results are shortly summarized in Fig.~\ref{Bezzi-cm}. 

As in the case of vegetative mycelia growth, 
different approaches are required to describe the microscopic details, in which
fluctuations and 
 stochasticity  play a major role.
Monte Carlo simulations are a possible choice, 
an alternative strategy is using cellular automata,
in which simple microscopic local rules (deterministic or probabilistic)
can produce complex patterns.

Cellular automata have also been used for studying pattern formation in
bacteria, in particular for {\it Streptomyces rutgersensis}~\cite{Mersi02}.
In this two-dimensional model the building block is the individual cell. 
A hypha
is composed of several cells, which do not move in space.  Cells
absorb food (energy) according to its local concentration,  convert
and store it internally and use the stored energy for metabolism,
growth and reproduction.
If the amount of stored energy lowers below a given threshold, the cells
start to starve. The model  assumes that this starvation phase triggers the
formation of aerial mycelium. 
When a cell reaches a given dimension and has accumulated sufficient
energy, it reproduces by binary division or gemmation.
Despite many approximations introduced,  
the resulting patterns of vegetative mycelium are very 
reminiscent of experimental patterns: 
at the beginning of the life of the colony 
(i.e., for the first generations) 
the hyphae are distributed randomly in space, but then, 
when food becomes scarce, they are selected 
to align with food gradient. 

Similar concentric rings patterns are also observed
in fungi, among them: {\sl Neurospora crassa}.
In Ref.~\cite{deutsch93}, Deutsch {\it et al.} present a cellular 
automaton  for modeling {\sl Neurospora crassa} growth patterns.
The two-dimensional patterns are reproduced using a
one-dimensional cellular automaton; but at each time step 
the one dimensional grid (called ''base space'' for the 
automaton) represents a circle of increasing radius
(so that the radial structure can be preserved).  
The model is capable of reproducing the morphologies of a 
single growing colony of {\it Neurospora Crassa},
but, due to the peculiar implementation of the radial 
symmetry, it is not applicable to the studies of interference 
patterns (similar to {\it S. rutgersensis} pattern  shown in Fig.~\ref{strepto-int}).

\section{Conclusions}

In filamentous organisms (e.g. fungi and actinomycetes), mother and daughter cells 
never detach: although they are separated by a septum,  
they remain physically connected. A filament of cells separated by
septa  is called an hypha, which is the building block of filamentous organisms. 
Hyphae are usually branched, 
and the complex of hyphae forming a fungal or actynomicete colony is called mycelium. 

The mycelial organization has deep consequences on growth of filamentous organisms, because active
growth is localized only at hyphal tips. Colony growth rate is not simply dictated
by hyphal tips growth, but also by the frequency  of formation of new tips --
branching frequency. Spatial organization and temporal
growth are intrinsically connected, and  the morphology of a colony is a key element to be taken
into account if we want to understand the growth of filamentous organisms.

Growth and mycelial pattern formation involve
events occurring at very different length scales, ranging from a colony scale (macroscopic)
to a single-cell scale (microscopic).
For example,  experimentally spatial organization is described by quantities like the angle between an hypha
and a related branch, the distance between two consecutive branches, the distance between two
consecutive septa, the time course of the total amount of branches, the frequency
of branches along an hypha, and so on. Evidently, these quantities
are measured at the microscopic scale, but the resulting
mycelial morphology is observed at the colony scale. 
On the other hand, the temporal dynamics of colony growth (e.g. colony radius growth rate)
is  a macroscopic  phenomenon, which is originated by tips growth, a single-cell event.

Mathematical models are needed to describe such complex phenomena that involve 
spatial and temporal dynamics, microscopic and macroscopic experimental
observations. Indeed, several models have been developed in the last thirty years to reproduce 
growth kinetics and mycelial patterns in filamentous organisms. 
Different biological questions  generate different type of modeling, that
 concur to draw a complete picture of filamentous organisms
growth and pattern formation. 
Such models differ for the
different organisms they apply to, as well as for the different scales of the experimental
observations they simulate.

In this comment we review a selection of models, organized according to the
 scale of the experimental observation from which the model originates.
We start with macroscopic models that reproduce colony growth rate experiments, 
then, moving down in length scale, to some
reaction diffusion models, that include more detailed phenomena like
septation and translocation, 
and finally we terminate with single hypha models,
that concern single hypha growth and branching. 
We also
review cellular automata models that implement microscopic 
interaction but at the same time they are able to reveal  macroscopic patterns.

The last step of the length scale should be the molecular level, but, to our knowledge, there are no models
that use the genetic data produced in the last twenty years to address growth kinetics
and pattern formation in filamentous organisms. Although genetic analysis revealed some of the
genes involved differentiation and the way they
are connected (i.e. the underlying gene networks),
 models based on these data are still missing. Such models have been produced for several basic processes
like the cell cycle, circadian rhythm and signal transduction pathways (for a review see \cite{Hasty01}).
 By using similar mathematical
analysis it should be possible to develop similar models to understand and modify in a predictable way
filamentous organisms.

\section*{Acknowledgements}
We thank F.~Bagnoli,    A.~Mengoni and G.~Mersi
for many helpful discussions. We also thank M.~Buiatti and O.~Coenen
for careful reading of the manuscript and suggesting many
improvements. We thank P.~Goatin for help in drawing the figures.
MB was supported by 
IST-2001-35271 grant (MB) and AC 
by DARPA-Biocomputation Program (AFRL \#F30602-02-0572).
Correspondence should be addressed to Michele Bezzi, Sony Computer Science Laboratory,
6 rue Amyot - 75005 Paris France. E-mail: michele@csl.sony.fr

\newpage
\section*{Figures}

\end{document}